\documentclass[10pt,journal,twocolumn]{IEEEtran}

\usepackage{amsmath,amssymb}
\usepackage{graphicx}
\usepackage{booktabs}
\usepackage{url}
\usepackage{tikz}
\usetikzlibrary{arrows.meta,positioning,fit,backgrounds}
\usepackage[hidelinks]{hyperref}
\usepackage{etoolbox}

\makeatletter
\g@addto@macro\normalsize{%
  \setlength\abovedisplayskip{4pt plus 2pt minus 1pt}%
  \setlength\belowdisplayskip{4pt plus 2pt minus 1pt}%
  \setlength\abovedisplayshortskip{1pt plus 2pt}%
  \setlength\belowdisplayshortskip{3pt plus 2pt minus 1pt}%
}
\makeatother
\begin{document}
\bstctlcite{IEEEtran:BSTcontrol}

\title{Semantic Split Inference for Remote\\
Modulation Recognition}

\author{\IEEEauthorblockN{Samer Lahoud, {\it Senior Member, IEEE} and  Kinda Khawam}
\thanks{Samer Lahoud is with the Faculty of Computer Science, Dalhousie University, Halifax, Canada (email: sml@dal.ca). Kinda Khawam is with LISN, Universit\'e Paris-Saclay, Gif-sur-Yvette, France (email: kinda.khawam@lisn.fr)}}

\maketitle

\begin{abstract}
Remote automatic modulation recognition balances sensing-node complexity,
reporting cost and accuracy. To address this trade-off, we propose
channel-aware semantic split inference: a sensing node sends a semantic report
over a noisy link and the edge server completes recognition. In our model,
split depth and report length are independent design variables, with
end-to-end training through the channel. We compare the resulting design
against basic split placements, which run inference at the edge server or at
the sensing node, and against a state-of-the-art collaborative scheme. We
assess sensing-node model size, computation, latency and energy against
recognition accuracy. We show that intermediate splits give the best accuracy--cost trade-off.
\end{abstract}


\section{Introduction}

Future wireless networks will require continuous awareness of increasingly
congested and heterogeneous spectrum. Automatic modulation recognition (AMR)
identifies the modulation format of an observed waveform and supports dynamic
spectrum access, interference analysis, and anomalous-signal
detection~\cite{zhang2026,oshea2018}. In large-scale monitoring systems,
resource-constrained sensing nodes observe the spectrum and an edge server
performs recognition. Forwarding raw in-phase and quadrature (I/Q) samples
preserves the sensed waveform but requires substantial uplink resources.
Running the complete classifier locally and reporting its output reduces
this load but requires the full model at every node.

This letter introduces \emph{channel-aware semantic split inference}, a
task-oriented communication framework for remote AMR~\cite{gunduz2023}. The
\emph{recognition network} is a neural classifier that maps a frame of I/Q
samples to a modulation label. We partition it between the sensing node and the
edge server at the \emph{split depth}, the network stage at which the report is
formed. There, the node maps its intermediate features to a compact
\emph{semantic report} and sends it over a noisy reporting channel; the edge
server completes inference~\cite{shao2020}. The report is optimized for
classification rather than waveform reconstruction. This yields two coupled
design questions: \emph{where in the network should the report be formed, and
how long should it be?} A shallow split limits the sensing-node model but
produces a weakly processed representation, whereas a deep split produces a
more task-specific representation at greater local cost. Report length sets
the number of channel uses and affects recognition accuracy.

Two limiting cases bound this choice. In one, the edge server performs almost
all of the inference and the node forwards a barely processed representation.
In the other, the node performs almost all of it and reports only a
decision. State-of-the-art collaborative-AMR schemes operate between
these limits but fix where the report is formed, so that report length has no
independent effect~\cite{dong2024}.

The contributions of this letter are threefold. First, we formulate remote AMR
with distinct sensing and reporting channels and evaluate both AWGN and flat
Rayleigh reporting links. Second, we vary split depth and report length
independently, including the allocation of a fixed length between semantic width
and temporal resolution. We characterize node model size, computation, measured
latency and energy, and accuracy. Third, under matched
conditions we compare the selected report against both basic split placements
and a state-of-the-art collaborative scheme. We show that intermediate splits
provide the best accuracy--cost trade-off at a matched reporting rate.

\section{Related Work and Positioning}\label{sec:related}
Deep learning has become the dominant approach to AMR: classifiers trained
directly on raw I/Q samples learn discriminative temporal representations and
outperform hand-designed feature pipelines~\cite{oshea2018}. The architectures
and signal representations that make this work are extensively
surveyed~\cite{zhang2026,peng2022}. Attention has also shifted from accuracy
alone to the cost of obtaining it, as in recognizers co-designed with hardware
accelerators for constrained platforms~\cite{wang2025mr}. This literature
nevertheless assumes that the observation and the classifier sit on the same
device.

That assumption fails in large-scale monitoring, where the design problem
extends from the classifier to the representation each node transmits.
Federated learning keeps observations local and exchanges model parameters
instead~\cite{wang2022fed}. This addresses privacy and training cost, but still
requires the complete model at every node at inference time. Split computing
takes the complementary route of partitioning a trained network between device
and server, so that the position of the cut and the size of the transmitted
tensor become the governing design choices~\cite{matsubara2023}. In this setting,
the link is usually treated as an error-free bit pipe. Two AMR schemes
follow this route and are closest to ours. C-AMC places a compression network at
the sensing node and a recurrent classifier at the edge
server~\cite{dong2024}. Its node network pools features to a fixed width
before the transmit projection, so the nominal report length can exceed the
intrinsic dimension of the representation it carries.
SplitAMC~\cite{park2023} instead splits a residual backbone after its first or
second stage and forwards the intermediate feature in full. This exposes no
report-length variable and requires at least eight times as many channel uses as
raw I/Q forwarding. SplitAMC is also the uncompressed baseline against
which~\cite{dong2024} reports. In neither case are
split depth and report length independent design variables, so the accuracy
obtained at a given reporting rate cannot be attributed to the split placement
alone. Separating the two is the starting point of this letter.

Our approach draws on task-oriented communication, which optimizes the
transmitted representation for a downstream task rather than for exact source
reconstruction~\cite{gunduz2023,chaccour2025}. Split inference realizes this
directly for classification: partitioning the recognition network lets the
node transmit a representation already shaped for the decision, with both
sides trained through a stochastic link~\cite{shao2020,wang2026}. The length
of that representation can itself be learned, for instance through an
information-bottleneck penalty that prunes uninformative
dimensions~\cite{shao2022}. We instead fix it by construction, which allows it
to be swept against split depth. Because the report is sent as
analog amplitudes rather than as a decoded bit stream, its accuracy degrades
gradually as the reporting channel worsens. Existing split-inference systems,
however, classify clean images~\cite{shao2020,wang2026,shao2022}. In remote AMR, the node input is
itself a noisy observation of an unknown waveform, so accuracy depends on both
a sensing and a reporting channel. The report can also retain temporal
structure, so its length can be divided between semantic width and token
count.

\section{System Model and Semantic Split Inference}\label{sec:system}

\paragraph{Sensing and reporting model}
A sensing node observes a complex-baseband frame of $N$ I/Q samples from a
signal with unknown modulation class $y\in\{1,\ldots,C\}$. The discrete-time
observation is modeled as
\begin{equation}\label{eq:sensing}
	r[n]
	=
	s_y[n]\,e^{j(2\pi\Delta f nT_s+\Theta)}
	+\eta_s[n],
	\qquad n=1,\ldots,N,
\end{equation}
where $s_y[n]$ is the waveform sample associated with class $y$, $\Delta f$
and $\Theta$ are residual carrier-frequency and phase offsets, $T_s$ is the
sampling period, and $\eta_s[n]\sim\mathcal{CN}(0,\sigma_s^2)$ is complex
additive white Gaussian noise. The sensing SNR
$\gamma_s=10\log_{10}(P_s/\sigma_s^2)$, with $P_s$ the average signal power,
characterizes the quality of the observed waveform. How each impairment
of~\eqref{eq:sensing} is realized is stated with~\eqref{eq:augment}. Stacking
the in-phase and quadrature components of $\mathbf{r}=[r[1],\ldots,r[N]]$
gives the real-valued network input
$\mathbf{x}=[\Re\{\mathbf{r}\};\Im\{\mathbf{r}\}]\in\mathbb{R}^{2\times N}$,
so that raw forwarding requires $L_x=2N$ real channel uses.

\begin{figure}[t]
\centering
\begin{tikzpicture}[
  >=Latex, font=\footnotesize,
  enc/.style={draw,rounded corners,fill=blue!10,minimum height=9mm,minimum width=17mm,align=center,inner sep=2pt},
  bn/.style={draw,rounded corners,fill=teal!16,minimum height=9mm,minimum width=17mm,align=center,inner sep=2pt},
  io/.style={font=\footnotesize},
  ch/.style={circle,draw,inner sep=1.6pt}]
\node[io] (x) {$\mathbf{x}$};
\node[enc,right=6mm of x] (Ek) {Encoder $E_k$\\[1pt]\scriptsize stages $1{:}k$};
\node[bn,right=7mm of Ek] (Qk) {TX $Q_k$\,+\,norm\\[1pt]\scriptsize $n_c{\times}T$};
\node[ch,right=7mm of Qk] (add) {$+$};
\node[above=3.5mm of add,align=center,font=\scriptsize] (n) {AWGN channel, $\gamma_r$\\[1pt]$\boldsymbol{\eta}_r\!\sim\!\mathcal{N}(\mathbf{0},\sigma_r^2\mathbf{I})$};
\node[bn,below=7mm of add] (Rk) {RX $R_k$};
\node[enc,left=7mm of Rk] (Gk) {Decoder $G_k$\\[1pt]\scriptsize stages $k{+}1{:}K$, head};
\node[io,left=7mm of Gk] (y) {$\widehat{y}$};
\draw[->] (x)--(Ek);
\draw[->] (Ek)-- node[above,inner sep=1.5pt]{$\mathbf{u}_k$} (Qk);
\draw[->] (Qk)-- node[above,inner sep=1.5pt]{$\mathbf{z}_k$} (add);
\draw[->] (n)--(add);
\draw[->] (add)-- node[right,inner sep=1.5pt]{$\widetilde{\mathbf{z}}_k$} (Rk);
\draw[->] (Rk)-- node[above,inner sep=1.5pt]{$\widehat{\mathbf{u}}_k$} (Gk);
\draw[->] (Gk)--(y);
\node[above=0.5mm of x,font=\scriptsize] {$\gamma_s$};
\begin{scope}[on background layer]
\node[draw,dashed,rounded corners,fill=black!3,fit=(Ek)(Qk),inner sep=2mm,label=above:{\scriptsize sensing node}] {};
\node[draw,dashed,rounded corners,fill=black!3,fit=(Gk)(Rk),inner sep=2mm,label=below:{\scriptsize edge server}] {};
\end{scope}
\end{tikzpicture}
\caption{Semantic split inference for remote AMR. The sensing node sends a
power-normalized report of $L_z=n_cT$ real channel uses over a noisy channel
(AWGN shown); the edge server completes inference.}
\label{fig:system}
\end{figure}
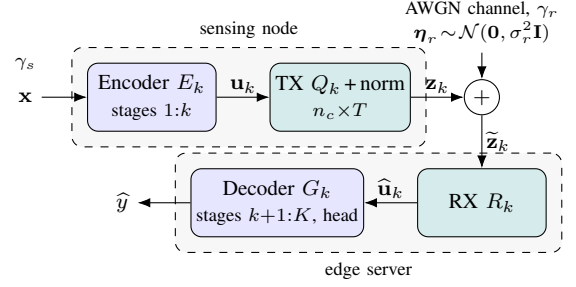

The overall pipeline is shown in Fig.~\ref{fig:system}. The recognition network
contains $K$ residual stages followed by a classification head. A split at
depth $k\in\{1,\ldots,K\}$ assigns the first $k$ stages to the sensing-node
encoder $E_k$ and the remainder to the edge server. A transmit interface $Q_k$
maps the encoder output $\mathbf{u}_k=E_k(\mathbf{x})$ to an unnormalized
report $\mathbf{b}_k=Q_k(\mathbf{u}_k)\in\mathbb{R}^{L_z}$, which is normalized
independently for each frame:
\begin{equation}\label{eq:pownorm}
	\mathbf{z}_k
	=
	\sqrt{L_z}\,
	\frac{\mathbf{b}_k}{\|\mathbf{b}_k\|_2},
	\qquad
	\frac{\|\mathbf{z}_k\|_2^2}{L_z}=1.
\end{equation}
The semantic report $\mathbf{z}_k$ thus has length $L_z$ real channel uses at
unit average power per use. As in analog joint source--channel
coding~\cite{deepjscc}, the normalization prevents unconstrained amplitude
scaling from artificially inflating the effective reporting SNR. One real
channel use is one real dimension of the transmitted baseband waveform, so
the report occupies $L_z/2$ complex symbols. The report is sent as uncoded
analog amplitudes, and reporting cost is therefore measured in channel uses
throughout. It is transmitted over an AWGN channel:
\begin{equation}\label{eq:report}
	\widetilde{\mathbf{z}}_k
	=
	\mathbf{z}_k+\boldsymbol{\eta}_r,
	\qquad
	\boldsymbol{\eta}_r
	\sim
	\mathcal{N}\!\left(
	\mathbf{0},
	\sigma_r^2\mathbf{I}_{L_z}
	\right),
	\qquad
	\sigma_r^2=10^{-\gamma_r/10},
\end{equation}
where $\gamma_r$ denotes the reporting SNR in decibels. The link is also
evaluated under flat Rayleigh block fading with perfect channel state
information at the edge server, specified in Sec.~\ref{sec:results}. Path
loss, shadowing and co-channel interference are not modeled separately. For a
fixed realization, each only shifts the average received power or the
effective noise floor, and is therefore absorbed into $\gamma_r$. For interference, this assumes that the interferer
is independent of the report and that the edge server performs no interference
cancellation. The reporting SNR $\gamma_r$ then reads as a
signal-to-interference-plus-noise ratio.

At the edge server, the receive interface $R_k$ maps the equalized report
$\widehat{\mathbf{z}}_k$ to the reconstructed feature tensor
$\widehat{\mathbf{u}}_k=R_k(\widehat{\mathbf{z}}_k)$, which has the dimensions
of $\mathbf{u}_k$. On the AWGN channel,
$\widehat{\mathbf{z}}_k=\widetilde{\mathbf{z}}_k$. The remaining residual
stages and the classification head $G_k$ then give
$\mathbf{p}_k=\operatorname{softmax}(G_k(\widehat{\mathbf{u}}_k))$ and
$\widehat{y}=\arg\max_c p_{k,c}$. When $k=K$, all residual stages are located
at the sensing node and $G_k$ contains only the classification head.

The split depth $k$ sets the model size and computation assigned to the sensing
node, whereas $L_z$ sets the reporting cost. Relative to raw forwarding, the
compression ratio is $\rho=L_x/L_z=2N/L_z$. The design $(k,n_c,T)$ therefore
trades sensing-node model size and computation
against reporting cost and recognition accuracy.

\paragraph{Backbone and semantic bottleneck}
We adopt a one-dimensional residual CNN based on the raw-I/Q AMR architecture
in~\cite{oshea2018}. The feature extractor contains $K$ residual stages of
width $w=64$. Each stage applies a $1\times1$ channel projection, two residual
convolutional units, and factor-two max pooling, so the encoder output at
split $k$ is $\mathbf{u}_k\in\mathbb{R}^{w\times L_k}$ with $L_k=N/2^k$. Global
average pooling and a two-layer classification head produce the $C$ output
logits.

The transmit interface applies a $1\times1$ projection reducing the feature
width from $w$ to $n_c$, and an adaptive average pooling $\mathcal{P}_T$
reducing the temporal length from $L_k$ to $T$:
\begin{equation}
	\mathbf{b}_k
	=
	\operatorname{vec}
	\left(
	\mathcal{P}_T
	\left(
	\operatorname{Conv}^{\mathrm{tx}}_{1\times1}
	(\mathbf{u}_k)
	\right)
	\right),
	\qquad
	L_z=n_cT,
\end{equation}
after which~\eqref{eq:pownorm} produces the transmitted report $\mathbf{z}_k$.
The parameters $n_c$ and $T$, which we call the semantic width and the token
count, allocate the report length: $T=1$ produces a single global
representation, whereas $T>1$ preserves a sequence of temporal feature tokens.
At the edge server, $R_k$ reshapes $\widehat{\mathbf{z}}_k$ into an
$n_c\times T$ tensor, a learned transposed convolution restores the temporal
length $L_k$, and a $1\times1$ receive projection restores the feature width
$w$, giving $\widehat{\mathbf{u}}_k$. A separate model is
initialized and trained for each evaluated $(k,n_c,T)$ configuration.

\paragraph{Basic split placements}
The two extreme split depths define the basic comparisons used in the
following. At $k=1$, the node model is as small as the parameterization allows,
but the report is formed from a barely processed representation. At $k=K$, the
entire backbone runs at the node, and no residual stage is left to process the
report after the channel.

\paragraph{Training}
The trainable modules $E_k$, $Q_k$, $R_k$, and $G_k$ are optimized end to end
through the stochastic reporting-channel layer by minimizing the cross-entropy
$\mathcal{L}_{\mathrm{CE}}=\mathbb{E}_{(\mathbf{x},y),\,\boldsymbol{\eta}_r}
[-\log p_{k,y}]$. For SNR-aware training,
the reporting SNR is redrawn for every mini-batch as
$\gamma_r\sim\mathcal{U}[\gamma_{\min},\gamma_{\max}]$, with an independent
noise realization applied to each frame. The channel of~\eqref{eq:report} enters
training as a parameter-free stochastic layer. Its Jacobian with respect to
$\mathbf{z}_k$ is the identity, so gradients reach $Q_k$ and $E_k$ unchanged and
no surrogate or straight-through estimator is required. We further apply random
phase rotation and circular time shifting:
\begin{equation}\label{eq:augment}
	\mathbf{x}'
	=
	\mathbf{R}(\varphi)\mathcal{S}_{\tau}(\mathbf{x}),
	\qquad
	\mathbf{R}(\varphi)
	=
	\begin{bmatrix}
		\cos\varphi & -\sin\varphi\\
		\sin\varphi & \cos\varphi
	\end{bmatrix},
\end{equation}
where $\varphi\sim\mathcal{U}[0,2\pi)$ and
$\tau\sim\mathcal{U}\{0,\ldots,N-1\}$; both transformations preserve the
modulation class and frame power. Each impairment of~\eqref{eq:sensing} is
exercised exactly once. The data generator realizes the sensing noise $\eta_s$
and the carrier-frequency offset $\Delta f$, drawing an independent center
frequency for every frame (Sec.~\ref{sec:results}). The residual phase is
supplied by the rotation, so that $\Theta=\varphi$. The circular shift
$\mathcal{S}_\tau$ has no counterpart in~\eqref{eq:sensing}: it represents the
unknown position of the signal within the observation window.

\section{Results}\label{sec:results}
We evaluate the proposed framework on a balanced, 16-class narrowband dataset
generated with TorchSig~\cite{torchsig}. The dataset spans ASK, PSK, QAM, FSK,
OFDM, and analog modulation families. Each frame contains $N=1024$
complex-baseband samples normalized to unit average power. During generation,
additive noise is applied at sensing SNR $\gamma_s$, while the center frequency
and bandwidth are randomized within the observation band.

At $k=3$, we sweep the full grid $n_c\in\{2,4,8,16\}$, $T\in\{1,2,4,8\}$,
which gives report lengths $L_z=n_cT$ from $2$ to $128$ real channel uses.
Accuracy does not increase beyond $L_z=64$, so results are shown for
$L_z\leq64$, that is, $\rho\geq32$. The split depth $k\in\{1,\ldots,5\}$ is
swept at $(n_c,T)\in\{(4,4),(8,4)\}$. Except in the last study of this
section, all models are trained over $\gamma_s\in[15,20]$~dB and evaluated at
$\gamma_s=20$~dB. In this regime the observation supports reliable
recognition, so accuracy differences between configurations reflect the split
and the reporting link. The reporting range $\gamma_r\in[0,20]$~dB extends from
a link whose noise power equals the report power to one that no longer limits
accuracy. Unless stated otherwise, the reporting SNR is sampled as
$\gamma_r\sim\mathcal{U}[0,20]$~dB during training. All learned schemes use
Adam with an initial learning rate of $10^{-3}$ reduced on plateau, a batch
size of $256$, and a 100-epoch training budget. The class-balanced training,
validation, and test sets contain $24\,000$, $4\,800$, and $6\,400$ frames,
respectively. Their partitions are fixed across runs. Accuracy results are
means over ten independent training runs. Where an uncertainty is given, it is
the sample standard deviation (SD) across those runs, in percentage points when
shown in parentheses.

The complete local classifier has $0.28$~M parameters, requires $203$~MFLOPs
per frame, and achieves a mean accuracy of $0.92$ (SD $0.01$) at
$\gamma_s=20$~dB. It serves as the no-reporting-channel reference.

\paragraph{Effect of split depth}
\begin{figure}[ht] \centering \includegraphics[width=\columnwidth]{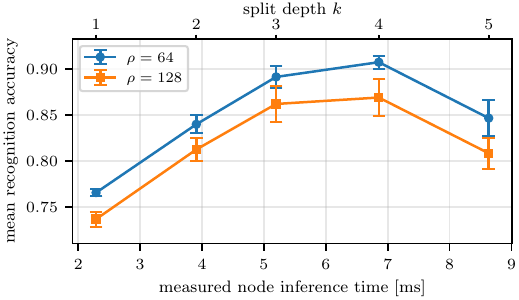} \caption{
		Mean recognition accuracy versus measured sensing-node inference time
		(Table~\ref{tab:jetson}) for split depths $k=1,\ldots,5$ and compression
		ratios $\rho\in\{64,128\}$ ($\gamma_s=\gamma_r=20$~dB). Error bars are
		$\pm1$ sample SD.}
\label{fig:splitdepth} \end{figure}
Fig.~\ref{fig:splitdepth} compares mean recognition accuracy across the $K=5$
split depths for $\rho\in\{64,128\}$. For both ratios, accuracy rises steeply
from $k=1$ to $k=3$, by about $12.5$ percentage points. This suggests that the
shallow encoder does not form a sufficiently discriminative representation
before compression. The subsequent gain from $k=3$ to $k=4$ is $0.7$ points at
$\rho=128$, within one SD, and $1.6$ points at $\rho=64$.

The lower-compression configuration $\rho=64$ is more accurate at every split.
Both follow the same increase--plateau--decrease pattern, identifying
$k\in\{3,4\}$ as the strongest region. From $k=4$ to $k=5$, accuracy decreases
by $6$ percentage points at both ratios, as no residual stage remains at the
edge server to process the noisy report. Recognition therefore depends on both
the representation formed before transmission and the edge-side capacity after
it.

FLOP counts alone do not predict the cost of a split on a deployed node, since
stages of comparable arithmetic cost differ in memory traffic and in
parallelism. We therefore measured the per-frame sensing-node cost of each
split on an NVIDIA Jetson AGX Orin at $30$~W, reported in
Table~\ref{tab:jetson}. At $k=3$, the node accounts for $90\%$ of the
FLOPs but only $58\%$ of the inference time. Energy follows the measured time
rather than the FLOP count: the node consumes $6.8$~mJ at that split, about
$61\%$ of the $11.1$~mJ of the complete local classifier. Choosing $k=3$ over
$k=4$ gives up the $0.7$ and $1.6$ points noted above, but saves $24\%$ of
node inference time and $27\%$ of node energy. We therefore adopt $k=3$ for
the experiments that follow, as the better accuracy--cost compromise.

\begin{table}[t]
	\caption{Sensing-node cost per frame on an NVIDIA Jetson AGX Orin ($30$~W,
		FP32, batch size one): median time over $1000$ inferences with
		interquartile range (IQR), and dynamic energy above idle. Percentages
		are of the local classifier ($203$~MFLOPs, $9.03$~ms, $11.1$~mJ).}
	\label{tab:jetson}
	\centering
	\setlength{\tabcolsep}{3pt}
	\small
	\begin{tabular}{cccccc}
		\toprule
		$k$ & MFLOPs & \% & time [ms] (IQR) & \% & energy [mJ] \\
		\midrule
		$1$ & $101$ & $50$  & $2.29$ $(2.0$--$2.5)$  & $25$ & $2.9$ \\
		$2$ & $156$ & $77$  & $3.91$ $(3.6$--$4.3)$  & $43$ & $4.5$ \\
		$3$ & $183$ & $90$  & $5.20$ $(5.0$--$5.7)$  & $58$ & $6.8$ \\
		$4$ & $196$ & $97$  & $6.86$ $(6.6$--$8.2)$  & $76$ & $9.3$ \\
		$5$ & $203$ & $100$ & $8.63$ $(8.2$--$10.4)$ & $96$ & $11.9$ \\
		\bottomrule
	\end{tabular}
\end{table}

\paragraph{Effect of semantic report length}
\begin{figure}[t]
	\centering
	\includegraphics[width=\columnwidth]{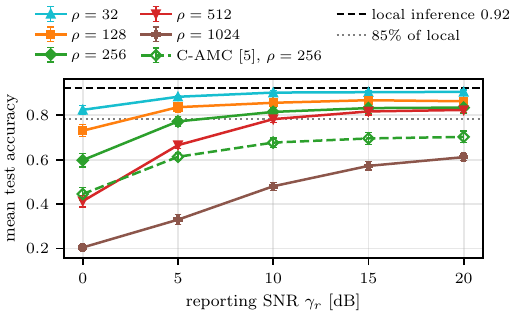}
	\caption{Mean test accuracy versus reporting SNR $\gamma_r$ at $k=3$,
		$\gamma_s=20$~dB, one curve per compression ratio $\rho$ at its
		validation-selected $(n_c,T)$. Open diamonds, dashed:
		C-AMC~\cite{dong2024} at the matched $\rho=256$. Error bars are $\pm1$
		sample SD.}
	\label{fig:ratefrontier}
\end{figure}

Fig.~\ref{fig:ratefrontier} reports, for each compression ratio at $k=3$, the
mean accuracy of the factorization $(n_c,T)$ selected on validation
accuracy. At $\gamma_r=20$~dB, accuracy degrades gradually with compression.
At $\rho=256$ it is $0.83$, which retains $90\%$ of the $0.92$ local-inference
accuracy. It remains above $0.80$ through $\rho=512$ and falls sharply to $0.61$
at $\rho=1024$.

This degradation depends strongly on the reporting channel: the shorter the
report, the more accuracy is lost as the channel worsens. Decreasing
$\gamma_r$ from $20$ to $0$~dB costs eight percentage points at
$\rho=32$ but $41$ points at $\rho=512$, from $0.82$ to $0.41$. At
$\gamma_r=0$~dB, accuracy has already fallen to $0.60$ at $\rho=256$. Shorter
reports leave fewer transmitted dimensions to withstand channel distortion.

Table~\ref{tab:alloc} isolates the allocation of a fixed report length
$L_z=n_cT$ between semantic width and temporal resolution. Allocation matters
most where the budget is scarcest: the spread across factorizations is
$13$ to $14$ percentage points at $L_z\leq16$ but only $2.7$ points at
$L_z=32$. The loss is driven by insufficient \emph{width} rather than by
insufficient temporal resolution. The $n_c=2$ configurations do not exceed
$0.75$ at any evaluated token count, whereas at $L_z\leq16$ every
configuration with $n_c\geq4$ lies within $1.3$ points of the best at the same
budget. A minimum semantic width is therefore necessary, above which the
division between width and time is not critical.

\begin{table}[t]
	\caption{Allocation of a fixed report length $L_z=n_cT$ at $k=3$,
		$\gamma_s=\gamma_r=20$~dB; the token count is $T=L_z/n_c$. Means over ten
		runs, sample SD in points in parentheses. Bold: best mean per length. A
		dash marks a factorization outside the swept grid.}
	\label{tab:alloc}
	\centering
	\setlength{\tabcolsep}{4pt}
	\small
	\begin{tabular}{ccccc}
		\toprule
		& \multicolumn{4}{c}{semantic width $n_c$} \\
		\cmidrule(lr){2-5}
		$L_z$ & $2$ & $4$ & $8$ & $16$ \\
		\midrule
		$8$  & $0.701$ $(1.4)$ & $\mathbf{0.844}$ $(1.4)$ & $0.835$ $(2.6)$ & --              \\
		$16$ & $0.745$ $(1.5)$ & $0.864$ $(0.5)$          & $\mathbf{0.876}$ $(1.7)$ & $0.863$ $(2.6)$ \\
		$32$ & --              & $0.869$ $(1.8)$          & $0.889$ $(0.8)$ & $\mathbf{0.896}$ $(1.4)$ \\
		\bottomrule
	\end{tabular}
\end{table}

\paragraph{Flat fading on the reporting link}
Under flat Rayleigh block fading the $L_z$ real uses are paired into $L_z/2$
complex symbols $z^{\mathrm{c}}_{k,i}$, received as
$\widetilde{z}^{\mathrm{c}}_{k,i}=h\,z^{\mathrm{c}}_{k,i}+\eta^{\mathrm{c}}_{i}$,
where a single gain $h\sim\mathcal{CN}(0,1)$ is drawn per frame and
$\eta^{\mathrm{c}}_{i}\sim\mathcal{CN}(0,2\sigma_r^2)$. We assume that the edge
server has perfect CSI. It applies the regularized linear equalizer
$\widehat{z}^{\mathrm{c}}_{k,i}=h^{*}\widetilde{z}^{\mathrm{c}}_{k,i}/(|h|^2+2\sigma_r^2)$,
and the real and imaginary parts of the result form $\widehat{\mathbf{z}}_k$.
The gain $|h|^2$ scales the instantaneous per-use SNR while leaving its average
unchanged. The
report is assumed to lie within one coherence time and bandwidth.
Table~\ref{tab:fading} retrains the same configurations through this link. At
$\gamma_r=20$~dB, the penalty relative to AWGN is only $1.2$ to $3.5$
percentage points, but it grows to $10.7$ points at $\gamma_r=0$~dB. Fading
thus amplifies the sensitivity to the reporting SNR: the accuracy lost between
$\gamma_r=20$ and $0$~dB is $1.9$ times larger under fading at $\rho=64$ and
$2.1$ times larger at $\rho=32$. At $\rho=1024$, the two channels lose accuracy
at nearly the same rate, because performance there is limited by the bottleneck
rather than by the link.

\begin{table}[t]
	\caption{Flat Rayleigh block fading against AWGN on the reporting link, at
		$k=3$, $\gamma_s=20$~dB, at the matched $(n_c,T)$ shown. Means over ten
		runs, sample SD in points in parentheses.}
	\label{tab:fading}
	\centering
	\setlength{\tabcolsep}{2pt}
	\small
	\begin{tabular}{cccccc}
		\toprule
		& & \multicolumn{2}{c}{$\gamma_r=20$~dB} & \multicolumn{2}{c}{$\gamma_r=0$~dB} \\
		\cmidrule(lr){3-4} \cmidrule(lr){5-6}
		$\rho$ & $(n_c,T)$ & AWGN & Rayleigh & AWGN & Rayleigh \\
		\midrule
		$32$   & $(16,4)$ & $0.892$ $(0.9)$ & $0.868$ $(3.0)$ & $0.834$ $(2.5)$ & $0.743$ $(1.4)$ \\
		$64$   & $(8,4)$  & $0.889$ $(0.8)$ & $0.877$ $(1.8)$ & $0.783$ $(2.0)$ & $0.676$ $(0.9)$ \\
		$256$  & $(8,1)$  & $0.835$ $(2.6)$ & $0.817$ $(1.3)$ & $0.598$ $(3.0)$ & $0.497$ $(1.7)$ \\
		$1024$ & $(2,1)$  & $0.612$ $(1.5)$ & $0.577$ $(1.7)$ & $0.204$ $(1.0)$ & $0.190$ $(0.9)$ \\
		\bottomrule
	\end{tabular}
\end{table}

\paragraph{Comparison with basic placements and the state of the art}
Table~\ref{tab:compare} sets the proposed split against the two basic placements
of Sec.~\ref{sec:system} and against C-AMC, all at the matched report length
$L_z=32$. All four schemes are trained on the same data, with the same
augmentation, training budget and reporting-SNR sampling. The three split rows
are the same backbone cut at different depths. C-AMC keeps its published
architecture. Its total size is similar, $0.27$~M parameters against our
$0.28$~M, but $93\%$ of them sit at the edge server. The C-AMC embedding is
placed under our per-frame power normalization and reporting channel. We report the
variant obtained without its published dropout stage, which we measured to
reduce inference-mode accuracy substantially at our frame length. This choice
favors the baseline.

The intermediate split is more accurate than both basic placements. Reporting
from the first stage has the lowest node cost but is $12.6$ points below
$k=3$. Reporting from the last stage requires $68\%$ more node parameters and
$66\%$ more node time than $k=3$ and remains $4.5$ points below it.

The node cost of C-AMC is lower still, $0.02$~M parameters against $0.16$~M
for the proposed split. The comparison is therefore a trade-off rather than a
dominance: C-AMC trails by $13.3$ points at this rate. Its accuracy is close to
that of our $k=1$ placement ($0.759$ against $0.766$), which also pairs a small
node model with a large edge model. The C-AMC gap is therefore comparable to
the $12.6$ points gained from $k=1$ to $k=3$, and much larger than the
difference between $k=3$ and $k=4$ on the plateau of Fig.~\ref{fig:splitdepth}.
In Fig.~\ref{fig:ratefrontier}, C-AMC at $\rho=256$ (dashed) lies below our
curve at every reporting SNR. At $\gamma_r=20$~dB, the gap stays
between $13.2$ and $16.0$ points at every report length from $L_z=2$ to $64$.

\begin{table}[t]
	\caption{Comparison at $L_z=32$ ($\rho=64$), $\gamma_s=\gamma_r=20$~dB.
		Split rows use $(n_c,T)=(8,4)$ and the runs of
		Fig.~\ref{fig:splitdepth}; independent runs of the proposed
		configuration give $0.889$ (Table~\ref{tab:alloc}). Means over ten
		runs, sample SD in points in parentheses.}
	\label{tab:compare}
	\centering
	\setlength{\tabcolsep}{4pt}
	\small
	\begin{tabular}{lcccc}
		\toprule
		& \multicolumn{2}{c}{Node} & Edge & \\
		\cmidrule(lr){2-3} \cmidrule(lr){4-4}
		Scheme & par. & MFLOPs & par. & Acc. \\
		\midrule
		$k=1$, min.\ depth & $0.05$\,M & $101$ & $0.24$\,M & $0.766$ $(0.4)$ \\
		$k=5$, max.\ depth & $0.27$\,M & $203$ & $0.01$\,M & $0.847$ $(2.0)$ \\
		C-AMC~\cite{dong2024} & $0.02$\,M & $36$  & $0.25$\,M & $0.759$ $(2.5)$ \\
		\midrule
		Proposed, $k=3$       & $0.16$\,M & $183$ & $0.12$\,M & $\mathbf{0.892}$ $(1.2)$ \\
		\bottomrule
	\end{tabular}
\end{table}

\paragraph{Sensing SNR and class-set size}
To evaluate lower sensing SNR, we retrained the $k=3$ split at $L_z=16$ and the
local classifier over $\gamma_s\in[-6,20]$~dB (Table~\ref{tab:sensing}). This
wider training range lowers the accuracy of both models at high $\gamma_s$. The
split retains about $90\%$ of the local accuracy down to $\gamma_s=5$~dB. The
loss caused by a poor reporting link shrinks as $\gamma_s$ falls, because the
observation itself becomes the limiting factor. At $\gamma_s=0$~dB, both models
are close to chance. We also repeated the comparison of
Table~\ref{tab:compare} with a reduced set of 8 classes and an expanded set of
24. With 24 classes, the ranking of the split depths is unchanged, and its
margins stay within one point of the 16-class values. With 8 classes, all
placements approach the local accuracy on a good reporting link, and the split
depth matters only when the link is poor.

\begin{table}[t]
	\caption{Mean accuracy versus sensing SNR at $k=3$, $L_z=16$, with both
		models trained over $\gamma_s\in[-6,20]$~dB. Sample SD in points in
		parentheses.}
	\label{tab:sensing}
	\centering
	\small
	\setlength{\tabcolsep}{5pt}
	\begin{tabular}{cccc}
		\toprule
		& & \multicolumn{2}{c}{Split} \\
		\cmidrule(lr){3-4}
		$\gamma_s$ [dB] & Local & $\gamma_r=20$~dB & $\gamma_r=0$~dB \\
		\midrule
		$20$ & $0.820$ $(2.5)$ & $0.721$ $(4.9)$ & $0.525$ $(3.8)$ \\
		$10$ & $0.581$ $(1.8)$ & $0.530$ $(5.5)$ & $0.368$ $(3.2)$ \\
		$5$  & $0.284$ $(3.5)$ & $0.264$ $(3.4)$ & $0.209$ $(2.8)$ \\
		$0$  & $0.119$ $(0.9)$ & $0.101$ $(1.0)$ & $0.099$ $(2.4)$ \\
		\bottomrule
	\end{tabular}
\end{table}

\section{Conclusion}

This letter decoupled network split depth from semantic report length in
remote AMR. It characterized the resulting trade-off among sensing-node
complexity, reporting cost and recognition accuracy over a noisy reporting
channel. Intermediate splits provided the best trade-off, and an eight-use
report retained $90\%$ of local-inference accuracy. At a matched reporting
rate, the intermediate split outperformed both basic placements and a
state-of-the-art collaborative scheme. Running the whole backbone at the sensing
node had the highest node cost and was still less accurate, because no
edge-side capacity remained to process the noisy report. The intermediate split kept its
advantage with 24 classes
and retained about $90\%$ of local accuracy at lower sensing SNR. Increasing
observation bandwidth raises encoder computation, while adaptive pooling keeps
the report length fixed by construction. Multi-antenna observation, faster
channel variation, frequency-selective reporting links and class-set mismatch
are left for future work.

\bibliographystyle{IEEEtran}
\bibliography{refs}

\end{document}